\documentclass[aps,pra,twocolumn,superscriptaddress,showpacs,10pt]{revtex4-1}
\usepackage{amsmath,amssymb,graphicx,color}
\usepackage{epstopdf}
\begin{document}

\title{Exploring global symmetry-breaking superradiant phase via phase competition
}

\author{Hai-Chao Li}
\altaffiliation{hcl2007@foxmail.com}
\affiliation{College of Physics and Electronic Science, Hubei Normal University, Huangshi 435002, China}

\author{Wen Huang}
\affiliation{School of Physics, Huazhong University of Science and Technology, Wuhan 430074, China}

\author{Wei Xiong}
\altaffiliation{xiongweiphys@wzu.edu.cn}
\affiliation{Department of Physics, Wenzhou University, Wenzhou 325035, China}

\begin{abstract}
Superradiant phase transitions play a fundamental role in understanding the mechanism of collective light-matter interaction at the quantum level. Here we investigate multiple superradiant phases and phase transitions with different symmetry-breaking patterns in a two-mode V-type Dicke model. Interestingly, we show that there exists a quadruple point where one normal phase, one global symmetry-breaking superradiant phase and two local symmetry-breaking superradiant phases meet. Such a global phase results from the phase competition between two local superradiant phases and can not occur in the standard $\Lambda$- and $\Xi$-type three-level configurations in quantum optics. Moreover, we exhibit a sequential first-order quantum phase transition from one local to the global again to the other local superradiant phase. Our study opens up a perspective of exploring multi-level quantum critical phenomena with global symmetry breaking.
\end{abstract}

\maketitle
The Dicke model~\cite{Dicke}, describing the collective interaction of $N$ two-level systems with a single-mode bosonic field, is a fundamental paradigm at the interface between quantum optics and condensed matter physics. Interestingly, this model predicts a second-order phase transition from a normal phase to a superradiant phase in the thermodynamics limit~\cite{Hepp,Wang,Emary1,Emary2}, characterized by macroscopic ground-state excitations of both the field and the two-level ensemble. At the same time, the ground state shows a twofold degeneracy in conjunction with a spontaneous $Z_2$ symmetry breaking. Such a phase transition originates from the singularity of quantum fluctuation at a quantum critical point and can occur at zero temperature when the light-matter coupling is scanned across the critical point. Except for providing valuable insights into many-body physics, superradiant phase transitions have potential applications in quantum information technologies~\cite{Lambert,Yang1,Yang2,Liuq}.

Cavity~\cite{Reiserer} and circuit~\cite{You,Blais} quantum electrodynamics (QED), opening up possibilities to tailor light-matter interaction at the quantum level, offer fascinating platforms for investigating superradiant phase transitions~\cite{Nataf,Viehmann,Bastidas,Buijsman,Lu}. In particular, superradiant phase transitions have already been simulated in many different quantum mechanical architectures, ranging from Bose-Einstein condensates~\cite{Baumann1} to thermal atoms~\cite{Zhiqiang}, trapped ions~\cite{Naini}, Fermi gases~\cite{Zhang}, free-space~\cite{Ferioli} and waveguide~\cite{Cardenas,Liedl} systems. However, all those works focus on the two-level Dicke model and the study of superradiant phase transitions remains a broad stage in three-level systems~\cite{Hayn,Ciuti,Baksic,Hayn1,Skulte,Kongkhambut,Lin,Fan}. More importantly, a three-level Dicke model introduces at least a new degree of freedom, thereby providing an important opportunity of exploring new quantum phases. From a theoretical perspective, it is indeed highly desirable to seek a novel phase for fundamental physics and quantum applications.

In this paper we show how a simple V-type Dicke model surprisingly can be used to generate a new superradiant phase accompanied by a global symmetry breaking and a fourfold degenerate ground state. We emphasize that this global superradiant phase can not exist in the standard $\Lambda$- and $\Xi$-type three-level Dicke models~\cite{Hayn}. In general, there are only two local symmetry-breaking superradiant phases and a highly unstable phase stemming from an imbalanced competition between these two local superradiant phases. Beyond expectation, such an unstable phase becomes a global symmetry-breaking superradiant phase under a balanced competition, which leads to the existence of a quadruple point where one normal phase and three superradiant phases meet. The global superradiant phase is superficially similar to the superradiant electromagnetic phase in the nonstandard two-level configuration~\cite{Baksic1} where each atom is coupled to both field quadratures of a boson mode. However, just because of the nonstandard atom-field interaction, the unusual model loses the universality in quantum optical platforms. Instead, our V-type Dicke model can be easily realized in various quantum systems, which can facilitate the experimental study. More importantly, due to the use of rich three-level configuration in our model, the phase competition can be introduced in a natural manner and this gives physical insight into the origin of the global superradiant phase. Thus, we can present a conversion from an unstable phase to a global superradiant phase by manipulating the phase competition. This work will help improve our understanding of the phase competition as well as the parity symmetry in quantum many-body physics.

\begin{figure}[htbp]
\centerline{
\includegraphics[width=0.7\columnwidth]{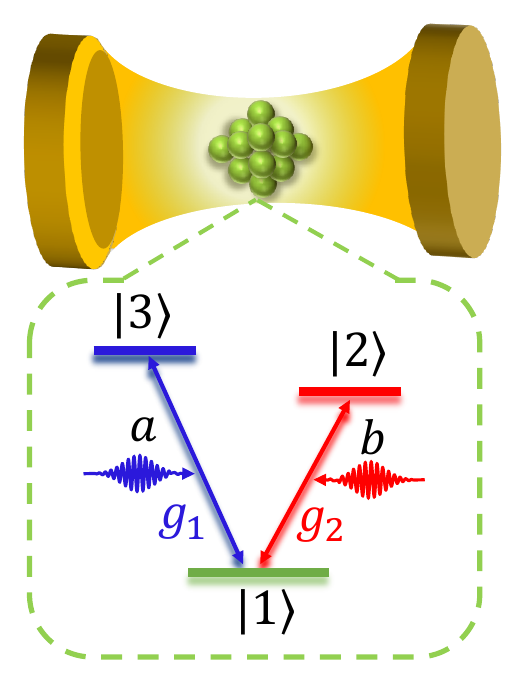}}
\caption{Schematic of a generalized V-type Dicke model with $N$ three-level systems identically coupled to two bosonic modes via strengths $g_{1}$ and $g_{2}$. Here levels $|1\rangle$ and $|3\rangle$ together with cavity mode $\emph{a}$ form a left branch while levels $|1\rangle$ and $|2\rangle$ as well as cavity mode $\emph{b}$ compose a right branch.
}
\label{fig.1}
\end{figure}

We consider a generalized Dicke model of $N$ three-level systems interacting with two bosonic modes, as shown in Fig.~\ref{fig.1}. Each system is described by a ground state $|1\rangle$ and two excited states $|2\rangle$ and $|3\rangle$ with frequencies $\omega_{i}$ ($i=1,2,3$). A cavity mode $\emph{a}$ with frequency $\omega_{a}$ is coupled to the $|1\rangle$$\leftrightarrow$$|3\rangle$ transition and another mode $\emph{b}$ with frequency $\omega_{b}$ is applied to the $|1\rangle$$\leftrightarrow$$|2\rangle$ transition, which constitutes a standard V-type configuration. The Hamiltonian of our model takes the form ($\hbar=1$)
\begin{align}\label{Hamil}
H&=\sum_{i=1}^{3}\omega_{i}J_{ii}+\sum_{k=a}^{b}\omega_{k}k^{\dag}k+\frac{g_{1}}{\sqrt{N}}(J_{13}+J_{31})(a^{\dag}+a)\nonumber \\
&\quad+\frac{g_{2}}{\sqrt{N}}(J_{12}+J_{21})(b^{\dag}+b),
\end{align}
where $J_{mn}=\sum_{j=1}^{N}|m\rangle^{j}\langle n|$ for $m,n=1,2,3$ are the collective operators and $g_{1,2}$ are the collective coupling strengths between the
cavity modes and the three-level ensemble. If $g_{1}=0$ ($g_{2}=0$), our three-level Dicke model reduces to a standard Dicke model which undergoes a superradiant phase transition at a critical value $g_{c2}=\sqrt{\omega_{b}\omega_{21}}/2$ ($g_{c1}=\sqrt{\omega_{a}\omega_{31}}/2$).

We now study the quantum criticality of the three-level Dicke model in the thermodynamic limit $N\rightarrow\infty$. It is convenient to introduce a multi-level Holstein-Primakoff transformation~\cite{Kurucz} to express the collective operators in terms of two bosonic modes $d_2$ and $d_3$. For a given reference state $|1\rangle$, the three-level boson mapping can be written as $J_{22}=d_2^{\dag}d_2$, $J_{33}=d_3^{\dag}d_3$, $J_{11}=N-d_2^{\dag}d_2-d_3^{\dag}d_3$, $J_{13}=\sqrt{N-d_2^{\dag}d_2-d_3^{\dag}d_3}d_3$, $J_{12}=\sqrt{N-d_2^{\dag}d_2-d_3^{\dag}d_3}d_2$, where the bosonic operators obey the commutation relations $[d_2, d_2^{\dag}]=1$ and $[d_3, d_3^{\dag}]=1$. Then we can utilize the mean-field approach and displace the bosonic operators in the following way: $a=\sqrt{N}\varphi_a+c_a$, $b=\sqrt{N}\varphi_b+c_b$, $d_{2,3}=\sqrt{N}\psi_{2,3}+e_{2,3}$. Here $\sqrt{N}\varphi_{a,b}$ and $\sqrt{N}\psi_{2,3}$ are the corresponding ground-state expectation values of the cavity modes and the collective modes. In this sense, the new operators $c_{a,b}$ and $e_{2,3}$ can be interpreted as the quantum fluctuations of the displaced bosonic modes. By expanding the Hamiltonian $H$ with the fluctuation operators and neglecting the terms with powers of $N$ in the denominator, we can have $H=Nh_0+\sqrt{N}h_1+h_2$, where $h_0$ is the scaled ground-state energy, $h_1$ is the linear terms in the fluctuation operators, and $h_2$ is the fluctuations around the ground state. We know that the order parameters $\varphi_{a,b}$ and $\psi_{2,3}$ can provide the relevant information for phases and phase transitions. In order to obtain these parameters, we set $h_1=0$ and give the conditional equations
\begin{align}\label{two}
\left(\omega_{21}+\frac{4g^{2}_{1}\psi^2_3}{\omega_a}+\frac{4g^{2}_{2}\psi^2_2}{\omega_b}-\frac{4g^{2}_{2}\psi^2_1}{\omega_b}\right)\psi_2&=0,\nonumber\\
\left(\omega_{31}+\frac{4g^{2}_{2}\psi^2_2}{\omega_b}+\frac{4g^{2}_{1}\psi^2_3}{\omega_a}-\frac{4g^{2}_{1}\psi^2_1}{\omega_a}\right)\psi_3&=0,
\end{align}
with $\varphi_a=-2g_1\psi_1\psi_3/\omega_a$, $\varphi_b=-2g_2\psi_1\psi_2/\omega_b$ and $\psi^2_1=1-\psi^2_2-\psi^2_3$. We find that there are four sets of solutions for Eq.~(\ref{two}). For example, a trivial solution is $\psi_2=\psi_3=0$ together with $\varphi_a=\varphi_b=0$, which corresponds to the normal phase. For convenience later, our three-level model can be divided into two different branches: the levels $|1\rangle$ and $|3\rangle$ together with the cavity mode $\emph{a}$ form a left branch while the levels $|1\rangle$ and $|2\rangle$ as well as the cavity mode $\emph{b}$ compose a right branch.

A nontrivial solution of Eq.~(\ref{two}) can be given by $\psi_3=\pm\sqrt{(1-\mu_l)/2}$, $\varphi_a=\mp{g_1}\sqrt{1-\mu^2_l}/\omega_a$ and $\psi_2=\varphi_b=0$ with $\mu_l=g^2_{c1}/g^2_1$. This solution is responsible for a superradiant phase in the left branch of the three-level Dicke model and it is only valid for $g_1\geq{g_{c1}}$. For clarity, we call this phase a left superradiant phase (as an analogy, a right superradiant phase would exist in the right branch). Interestingly, owing to the left superradiant regime, there is a second quantum critical point in the right branch besides the fixed $g_{c2}$. To be specific, by inserting the above solution into the expression $h_2$, we find that $h_2$ can be divided into two independent subexpressions. One of them is related to the left superradiant phase, and the other one describing a renormalized Hamiltonian for the right branch is written as
\begin{align}\label{three}
h_r&=\omega_bb^\dag b+\widetilde{\omega}_{21}d^\dag_2d_2+\widetilde{g}_2(d_2^{\dag}+d_2)(b^{\dag}+b),
\end{align}
with the coefficients $\widetilde{\omega}_{21}=\omega_{21}+\omega_{31}(1-\mu_l)/(2\mu_l)$ and $\widetilde{g}_2=g_{2}\sqrt{(1+\mu_l)/2}$. Based on the Hamiltonian $h_r$, a new quantum critical point can be easily given by $\widetilde{g}_{c2}=\frac{1}{2}\sqrt{1/(1+\mu_l)}\sqrt{2\omega_{21}\omega_b+\omega_{31}\omega_b(1-\mu_l)/\mu_l}$. Here the right branch is not macroscopically occupied in the coupling $g_2\leq\widetilde{g}_{c2}$, as demonstrated by $\psi_2=\varphi_b=0$. In physics, the new critical point results from the common level $|1\rangle$ involved in the two branches, which can be explained using a simple physical picture. When $g_1\leq{g}_{c1}$, we have $\psi_3=\varphi_a=0$ and no left superradiant phase occurs. In this case, all three-level systems occupy their respective level $|1\rangle$ with $\psi_1=1$ and so the fixed critical point ${g}_{c2}$ is obtained in the right branch. However, when $g_1\geq{g}_{c1}$, the left branch acquires macroscopic excitations. At the same time, the occupation of the level $|1\rangle$ is modified as $\psi^2_1=(1+\mu_l)/2$ and the energy shift $\omega_{31}(1-\mu_l)/(2\mu_l)$ is transferred to the mode $d_{2}$ by the level $|1\rangle$, which lead to the renormalized Hamiltonian $h_r$. In the left superradiant phase, the critical point $\widetilde{g}_{c2}$ depends on the coupling strength $g_1$ via $\mu_l$. For $g_1=g_{c1}$, $\widetilde{g}_{c2}$ reduces to $g_{c2}$. As $g_1$ is increased, $\widetilde{g}_{c2}$ increases, until it eventually tends to $g_{1}\sqrt{\omega_b/\omega_{a}}$ in the $g_1\rightarrow\infty$ limit.

\begin{figure}[htbp]
\centerline{
\includegraphics[width=0.98\columnwidth]{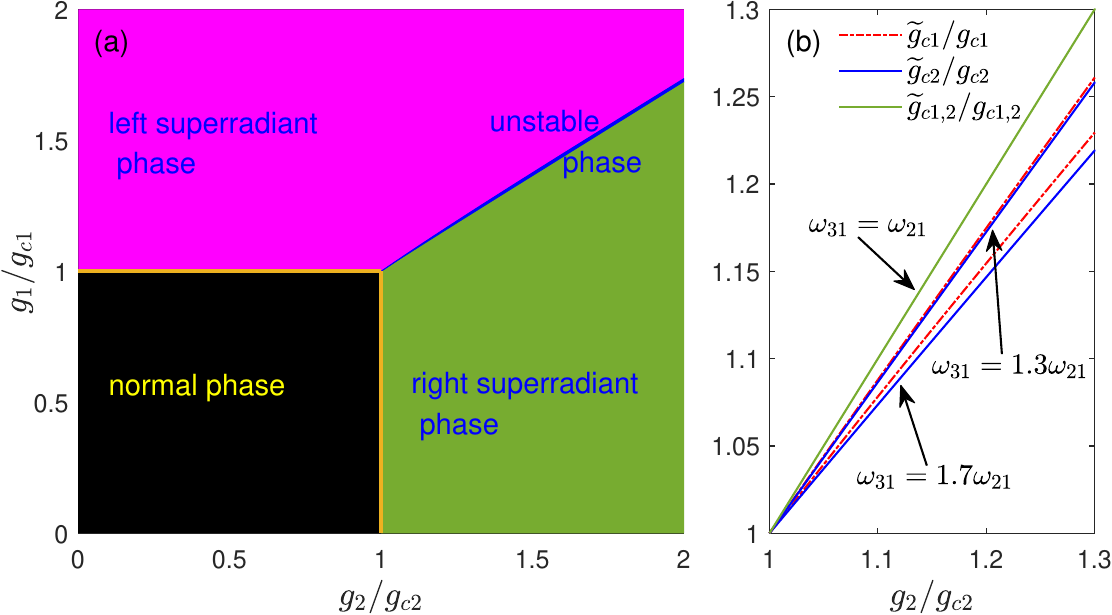}}
\caption{(a) Phase diagram in the two dimensional ($g_{2}$, $g_{1}$) plane at $\omega_{31}=1.7\omega_{21}$. The blue region encompassed by two parameterized phase boundaries $\widetilde{g}_{c1}$ and $\widetilde{g}_{c2}$ indicates an unstable phase deriving from an imbalanced competition between the left and right superradiant phases. (b) Frequency-dependent overlapping region. As the ratio $\omega_{31}/\omega_{21}$ decreases, the region gradually lessens and eventually reduces to a diagonal line at $\omega_{31}=\omega_{21}$.
}
\label{fig.2}
\end{figure}

Similarly, there is a right superradiant phase described by another nontrivial solution $\psi_2=\pm\sqrt{(1-\mu_r)/2}$, $\varphi_b=\mp{g_2}\sqrt{1-\mu^2_r}/\omega_b$ and $\psi_3=\varphi_a=0$ with $\mu_r=g^2_{c2}/g^2_2$. In the right superradiant phase, the Hamiltonian $h_l$ responsible for the left branch is corrected as
\begin{align}\label{four}
h_l&=\omega_aa^\dag a+\widetilde{\omega}_{31}d^\dag_3d_3+\widetilde{g}_1(d_3^{\dag}+d_3)(a^{\dag}+a),
\end{align}
which brings about the fourth quantum critical point $\widetilde{g}_{c1}=\frac{1}{2}\sqrt{1/(1+\mu_r)}\sqrt{2\omega_{31}\omega_a+\omega_{21}\omega_a(1-\mu_r)/\mu_r}$. The parameters in Eq.~(\ref{four}) are $\widetilde{\omega}_{31}=\omega_{31}+\omega_{21}(1-\mu_r)/(2\mu_r)$ and $\widetilde{g}_1=g_{1}\sqrt{(1+\mu_r)/2}$. As a result, we also demonstrate two critical points in the left branch and $\widetilde{g}_{c1}$ relaxes back to ${g}_{c1}$ at $g_2=g_{c2}$. Clearly, the left branch in the right superradiant phase can show a similar quantum criticality with the right branch in the left superradiant phase.

We further explore simultaneous macroscopic excitations for the two branches, which would require $\psi_2\neq0$ and $\psi_3\neq0$ in Eq.~(\ref{two}). The corresponding phase is named as a left-right superradiant phase. A general solution for this superradiant phase can be described by
\begin{align}\label{five}
(\alpha-\beta)^2\psi^2_2&=2\alpha(\omega_{21}-\beta)-(\omega_{31}-\alpha)(\alpha+\beta),\nonumber \\
(\alpha-\beta)^2\psi^2_3&=2\beta(\omega_{31}-\alpha)-(\omega_{21}-\beta)(\alpha+\beta),
\end{align}
with $\alpha=\omega_{31}/\mu_l$ and $\beta=\omega_{21}/\mu_r$. When $\alpha\neq\beta$, this equation yields two subtle relations $g_1<\widetilde{g}_{c1}$ and $g_2<\widetilde{g}_{c2}$. These inequalities show a contradictory physics because they hold both in the left-right superradiant phase and in the right (left) superradiant phase where the left (right) branch is only microscopically excited under $g_1<\widetilde{g}_{c1}$ ($g_2<\widetilde{g}_{c2}$). Such a contradiction indicates that the so-called left-right superradiant phase is physically unstable or even nonexistent.

According to the above discussion, we plot a phase diagram in Fig.~\ref{fig.2}(a). Theoretically, this diagram should display three well-defined phases: a normal phase, a left superradiant phase and a right superradiant phase, where the parameterized phase boundary for the left (right) superradiant phase is governed by $\widetilde{g}_{c2}$ ($\widetilde{g}_{c1}$) versus the coupling $g_1$ ($g_2$). However, an unexpected overlapping region between the two superradiant phases emerges, as depicted in Fig.~\ref{fig.2}(a) with blue. Obviously, this region denotes just the unstable left-right superradiant phase. In physics, such an unstable phase is due to an imbalanced competition between the left and right superradiant phases. Note that the overlapping region is not fixed and it decreases as the frequency ratio $\omega_{31}/\omega_{21}$ diminishes, as shown in Fig.~\ref{fig.2}(b). Ultimately, it becomes a diagonal line when $\omega_{31}=\omega_{21}$. But the line would be excluded from the unstable region because every point in the line satisfies the expression $\alpha=\beta$. This exception plays a crucial role in the stable left-right superradiant phase, as demonstrated below.

\begin{figure}[htbp]
\centerline{
\includegraphics[width=0.98\columnwidth]{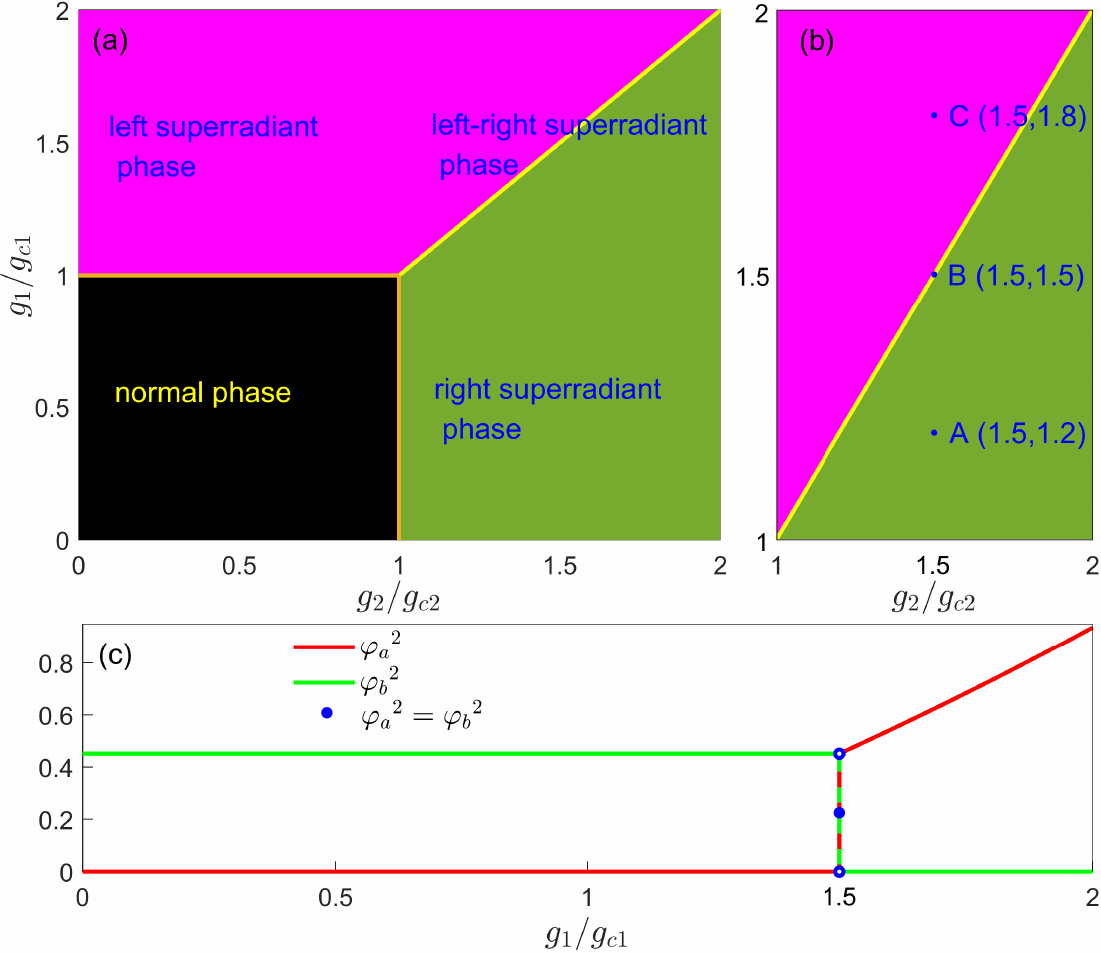}}
\caption{(a) Phase diagram including a stable left-right superradiant phase with a fourfold degenerate ground state at $\omega_{31}=\omega_{21}$. This diagram shows a quadruple point where one normal phase and three superradiant phases meet at $g_1=g_{c1}$ and $g_2=g_{c2}$. (b) An example showing a sequential first-order superradiant phase transition. (c) The plotted order parameters ${\varphi_a}^2$ and ${\varphi_b}^2$ with $g_2/g_{c2}=1.5$.
}
\label{fig.3}
\end{figure}

Interestingly enough, a particular solution exists at $\alpha=\beta$ making the system fully degenerate, despite the fact that a general left-right superradiant phase is removed. In this condition, we have $\psi_3=\pm\sqrt{1-\mu_l}/2$, $\varphi_a=\mp{g_1}\sqrt{1-\mu^2_l}/(\sqrt{2}\omega_a)$ in the left branch and simultaneously $\psi_2=\pm\sqrt{1-\mu_r}/2$, $\varphi_b=\mp{g_2}\sqrt{1-\mu^2_r}/(\sqrt{2}\omega_b)$ in the right branch. This phase displays a symmetric structure in the two branches because of $\omega_{31}=\omega_{21}$, $\omega_a=\omega_b$ and $g_1=g_2$ and thus can be interpreted as a result of balanced competition between these two branches. Obviously, such a left-right superradiant phase has a fourfold degenerate ground state by simply analysing the signs of the order parameters. Moreover, the parameterized critical curves $\widetilde{g}_{c1}$ and $\widetilde{g}_{c2}$ merge into a diagonal line, as seen in Fig.~\ref{fig.2}(b). The diagonal line not only represents a stable left-right superradiant phase, but also is an exact phase boundary between the left and right superradiant phases, as shown in Fig.~\ref{fig.3}(a). From the picture, we find a quadruple point where the four phases meet at $g_1=g_{c1}$ and $g_2=g_{c2}$.
In addition, the diagonal line defines a first-order phase transition line where the order parameters have an abrupt change. We stress that such phase transition is observable under appropriate parameters. For instance, by setting $g_2$ to a fixed value and enhancing $g_1$ gradually in Fig.~\ref{fig.3}(b), a sequential transition from the right to the left-right again to the left superradiant phase can be predicted. This prediction is verified by Fig.~\ref{fig.3}(c) where the order parameters ${\varphi_a}^2$ and ${\varphi_b}^2$ show a jump across the phase transition point. Note that the quadruple point along with two brown curves forms a second-order phase transition line.

\begin{figure}[htbp]
\centerline{
\includegraphics[width=0.98\columnwidth]{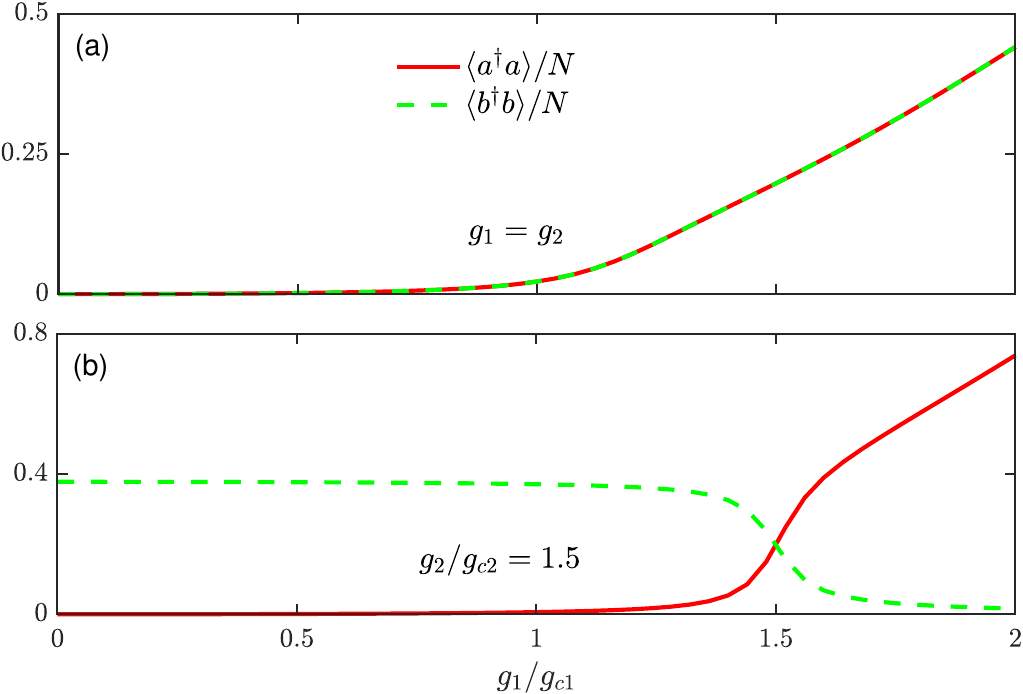}}
\caption{The scaled ground-state expectation values $\langle a^\dag a\rangle/N$ and $\langle b^\dag b\rangle/N$ at $\omega_{31}=\omega_{21}$.
}
\label{fig.4}
\end{figure}

To support our theoretical results in the thermodynamic limit, we numerically investigate the generalized V-type Dicke model at finite $N=10$ in Fig.~\ref{fig.4} where the scaled ground-state expectation values $\langle a^\dag a\rangle/N$ and $\langle b^\dag b\rangle/N$ are plotted at $\omega_{31}=\omega_{21}$. Obviously, the existence of the left-right superradiant phase can be confirmed in Fig.~\ref{fig.4}(a) with identical macroscopic excitations. The sequential phase transition shown in Fig.~\ref{fig.3}(c) is also checked by the tendency in Fig.~\ref{fig.4}(b) where the difference around the critical point $g_1=1.5g_{c1}$ results from the finite $N$.

The spontaneous $Z_2$ symmetry breaking is an important character of superradiant phase transitions. In analogy to the standard Dicke Hamiltonian, we can use a parity operator to describe a symmetry~\cite{Felicetti} in our model. We define two local parity operators $\prod_l=\textrm{exp}[i\pi(a^{\dag}a+J_{33})]$ in the left branch and $\prod_r=\textrm{exp}[i\pi(b^{\dag}b+J_{22})]$ in the right branch. The operators $\prod_l$ and $\prod_r$ commute with the Hamiltonian $H$ in Eq.~(\ref{Hamil}), which can be certified by the following transformations $\prod_l:(a,J_{31})\rightarrow\prod_l(a,J_{31})\prod^{\dag}_l=(-a,-J_{31})$ and $\prod_r:(b,J_{21})\rightarrow\prod_r(b,J_{21})\prod^{\dag}_r=(-b,-J_{21})$. Meanwhile, a global parity operator can be written as a product of two local parity operators $\prod_g=\prod_l\prod_r$. It is obvious that our model has a conserved parity symmetry in the normal phase while the left (right) superradiant phase breaks the local $\prod_l$ ($\prod_r$) symmetry. In the degenerate condition, the existence of the left-right superradiant phase leads to the global symmetry breaking where the ground state becomes fourfold degenerate. As a result, the left-right superradiant phase is regarded as a global phase and the left (right) superradiant phase is a local phase.

We have explored superradiant phase transitions by means of a standard mean-field theory in a generalized three-level Dicke model. We present a rich phase diagram of superradiant phases with different symmetry breakings and phase transitions with exact first- and second-order boundaries. In particular, a balanced competition between two local superradiant phases gives rise to a global superradiant phase, triggering a quadruple point in which one normal phase and three superradiant phases can coexist. Given an impressive ongoing progress of manipulating light-matter interaction at the quantum level, our model can be used to study quantum critical phenomena in a wide range of physical systems, such as natural atoms, trapped ions, and superconducting circuits. From a broader viewpoint, our work offers the prospect of exploring global phase-transition-related physics where all motional degrees of freedom of a quantum many-body system are macroscopically excited.

This work was partially supported by the National Natural Science Foundation of China under Grant No. 11904201.

\end{document}